\documentclass[aps,prd,twocolumn,showpacs,floats]{revtex4}

\usepackage{graphicx}
\usepackage{bm}

\newcommand{\eqref}[1]{(\ref{#1})}
\newcommand{\khat}{\hat k}
\newcommand{\zhat}{\hat z}
\newcommand{\ahat}{\hat a}
\newcommand{\bhat}{\hat b}
\newcommand{\chat}{\hat c}
\newcommand{\dhat}{\hat d}
\newcommand{\ehat}{\hat e}

\begin{document}

\title{Toward Stable 3D Numerical Evolutions of Black-Hole Spacetimes}

\author{Mark A. Scheel${}^1$, Lawrence E. Kidder${}^2$, Lee Lindblom${}^1$,
Harald P. Pfeiffer${}^2$, and Saul A. Teukolsky${}^2$}

\affiliation{${}^1$ Theoretical Astrophysics 130-33, California
Institute of Technology, Pasadena, CA 91125} 
\affiliation{${}^2$ Center for Radiophysics and Space Research,
Cornell University, Ithaca, New York 14853}

\date{\today}

\begin{abstract}
Three dimensional (3D) numerical evolutions of static black holes
with excision are presented. These evolutions extend to about $8000M$,
where $M$ is the mass of the black hole.  This degree of stability is
achieved by using growth-rate estimates to guide the fine tuning of
the parameters in a multi-parameter family of symmetric hyperbolic
representations of the Einstein evolution equations.  These evolutions
were performed using a fixed gauge in order to separate the intrinsic
stability of the evolution equations from the effects of stability-enhancing
gauge choices.
\end{abstract}

\pacs{04.25.Dm, 04.20.Ex, 02.60.Cb, 02.30.Mv}

\maketitle

Recent studies have documented the fact that constraint-violating
instabilities are a common (if not universal) feature of solutions to
the Einstein evolution
equations~\cite{Kidder2001,Laguna2002,Lindblom2002}.  Initial data
with small numerical errors on some initial Cauchy surface will
typically evolve to a solution in which the constraints grow
exponentially with time.  Black-hole spacetimes that are evolved in
full 3D (without symmetry) with a fixed gauge using one of the
``standard'' formulations of the evolution equations
(e.g. ADM\cite{ADM,york79} or BSSN\cite{shibata95,baumgarte99}) have
instabilities of this type that become unphysical (e.g. because the
constraints become large) on a timescale of about
100$M$~\cite{Yo2002,Alcubierre2000b}, where $M$ is the mass of the
black hole.  Several studies have shown that changing the evolution
equations by adding multiples of the constraints and by changing the
dynamical fields can have a significant effect on the growth rate of
these constraint-violating
instabilities~\cite{Kidder2001,Laguna2002,Lindblom2002}.  Such a
re-formulation of the BSSN evolution equations has allowed full 3D
evolutions with fixed gauge to persist for about
1400$M$~\cite{Shoemaker2002}.  The duration of black hole evolutions
has also been extended considerably, apparently indefinitely in some
cases, by imposing symmetries, e.g. octant, on the
solutions~\cite{Alcubierre2001} or by using an appropriate dynamical
gauge~\cite{Yo2002,Alcubierre2002}.

We present new results for evolving isolated static black holes using
a multi-parameter family of symmetric hyperbolic representations of
the Einstein evolution equations~\cite{Kidder2001}. For the optimal
case our evolutions extend to about $8000M$.  We focus on the question
of how the evolution equations themselves affect stability, and
therefore we use a fixed gauge~\footnote{We use a fixed {\it
densitized\/} lapse and fixed shift. For the evolution equations studied
here fixing the lapse itself violates hyperbolicity, which is
inconsistent with the way our code imposes boundary conditions.}  and
do not impose any symmetries on the solutions.  The fine tuning needed
to achieve optimal stability for evolving a single black hole requires
a special choice of the parameters in our representation of the
evolution equations, but does not require any fine tuning of our
numerical methods.  Thus we expect that any numerically stable
evolution code that solves this same system of equations with the same
initial data and boundary conditions will exhibit the same behavior we
find here.

We study the evolution of black-hole spacetimes using a particular
12-parameter family of representations of the Einstein evolution
equations~\cite{Kidder2001}.  This family is derived from the standard
3+1 ``ADM'' form of the equations by introducing five parameters
$\{\gamma,\zeta,\eta,\chi,\sigma\}$ that densitize the lapse function
and add multiples of the constraints to the evolution equations, and
seven additional parameters
$\{\zhat,\khat,\ahat,\bhat,\chat,\dhat,\ehat\}$ that re-define the set
of dynamical fields.  The details of the resulting evolution equations
and the precise definitions of these various parameters are explained
at length elsewhere~\cite{Kidder2001,Lindblom2002}, and will not be
repeated here.  It has been shown that a 9-parameter subfamily of
these representations consists of strongly hyperbolic evolution
equations in which all of the characteristic speeds of the system
(relative to the hypersurface-normal observers) have only the physical
values: $\{0,\pm 1\}$~\cite{Kidder2001}.  It has also been shown that
the evolution equations for an open subset of this 9-parameter family,
in particular those representations with $-\frac{5}{3}<\zeta<0$, are
symmetric hyperbolic~\cite{Lindblom2002}.  Our numerical analysis here
is confined to this 9-parameter family of symmetric hyperbolic
representations of the Einstein evolution equations having physical
characteristic speeds.

Here we analyze the numerical
evolution of initial data that represents a single isolated static
black hole.  For initial data we use a $t={\rm constant}$
slice of the Schwarzschild geometry written in Painlev\'e-Gullstrand
coordinates~\cite{Martel2000},
\begin{equation}
ds^2 = -dt^2 + \left(dr+\sqrt{\frac{2M}{r}}dt\right)^2+r^2 d\Omega^2,
\label{e:pgmetric}
\end{equation}
(where $d\Omega^2$ is the standard metric on the unit sphere), plus
small perturbations that are added by hand.  By explicitly inserting
the same perturbations for all numerical resolutions, we are able to
test convergence; this would not be the case if instead we allowed the
perturbations to arise from machine roundoff error.  The exact form of
the perturbations is unimportant; it does not affect either the
asymptotic growth rate of the unstable mode or its spatial dependence.

We also fix the gauge for these evolutions (not just at the initial
time but throughout the evolution) by setting the densitized lapse and
the shift to those of Eq.~\eqref{e:pgmetric}.  Fixing the gauge in
this way is known to be less stable than using a carefully selected
dynamically determined gauge~\cite{Yo2002,Alcubierre2000b}.  However,
our purpose here is to study the intrinsic stability of the evolution
equations, so we choose to fix the gauge in this non-optimal way in
order to isolate and emphasize this instability.

The evolution equations are solved here using a pseudospectral
collocation method (see \cite{Kidder2001,Kidder2000a,Kidder2002a} for
further details on the implementation) on a 3D spherical shell
extending (typically) from $r=1.9M$ to $r=11.9M$.  This code utilizes
the method of lines; the time integration is performed using a
fourth-order Runge-Kutta algorithm.  Although we use spherical
coordinates, our fundamental variables are the Cartesian components of
the various fields. The inner boundary lies inside the event horizon;
at this boundary all the characteristic curves are directed out of the
domain (into the black hole), so no boundary condition is required
there and none is imposed (``horizon excision'').  At the outer
boundary we require that all ingoing characteristic fields be
time-independent, but we allow all outgoing characteristic fields to
propagate freely.

\begin{figure}
\begin{center}
\includegraphics[width=3.0in]{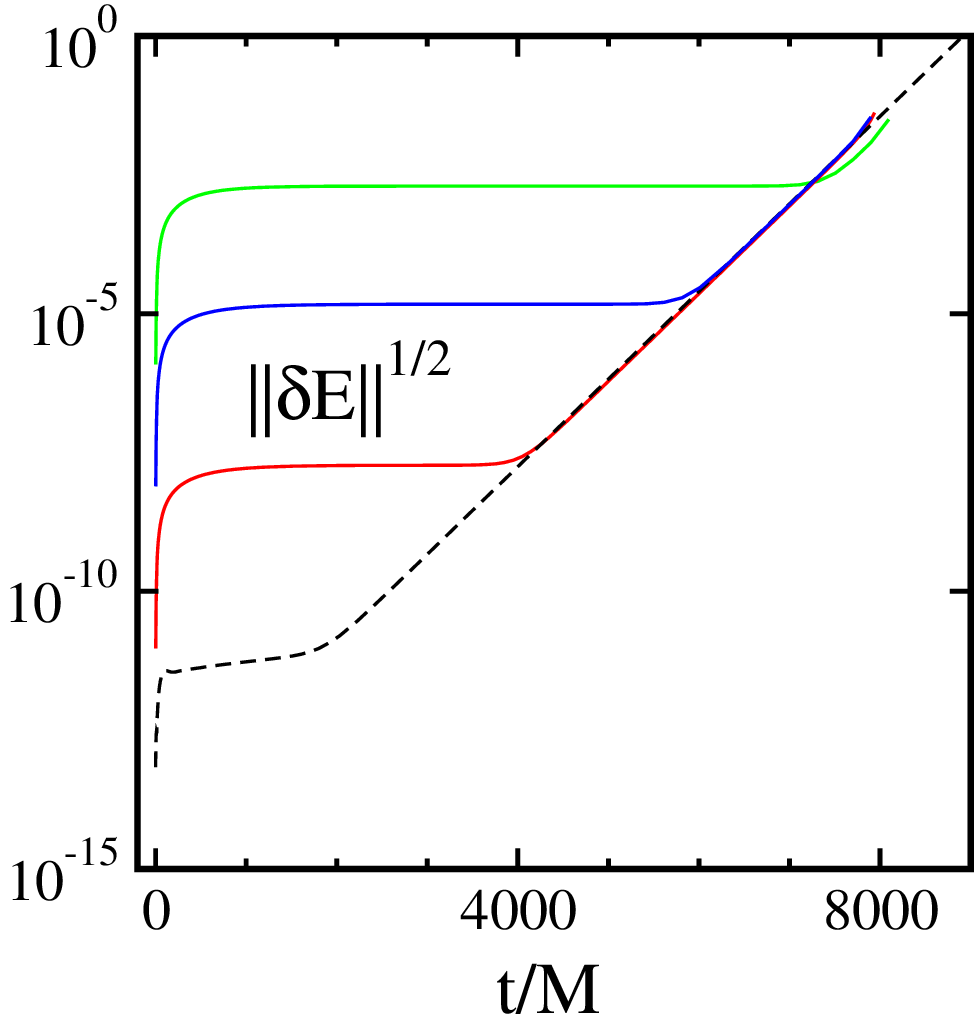}
\end{center}
\caption{Energy norm $||\delta E||{}^{1/2}$
(per unit volume) for the most stable set of evolution parameters.
Solid curves are $||\delta E||{}^{1/2}$ from the full non-linear evolution
code, and the dashed curve is from a linearized version of the code.}
\label{f:Enorms}
\end{figure}

Recent analytical work~\cite{Lindblom2002} has shown that the growth
rates of the constraint-violating instabilities for the
Painlev\'e-Gullstrand form of the Schwarzschild geometry depend on
just three of the nine parameters that specify the evolution
equations, $\{\gamma,\zeta,\zhat\}$.  We confine our study here to the
dependence of this instability on the two parameters
$\{\gamma,\zhat\}$~\footnote{The analytical
estimates~\cite{Lindblom2002} suggest that varying the third parameter
$\zeta$ should not result in significantly increased stability.},
and we fix the
remaining parameters to the values that define System III of
Ref.~\cite{Kidder2001}.

\begin{figure}
\begin{center}
\includegraphics[width=3.0in]{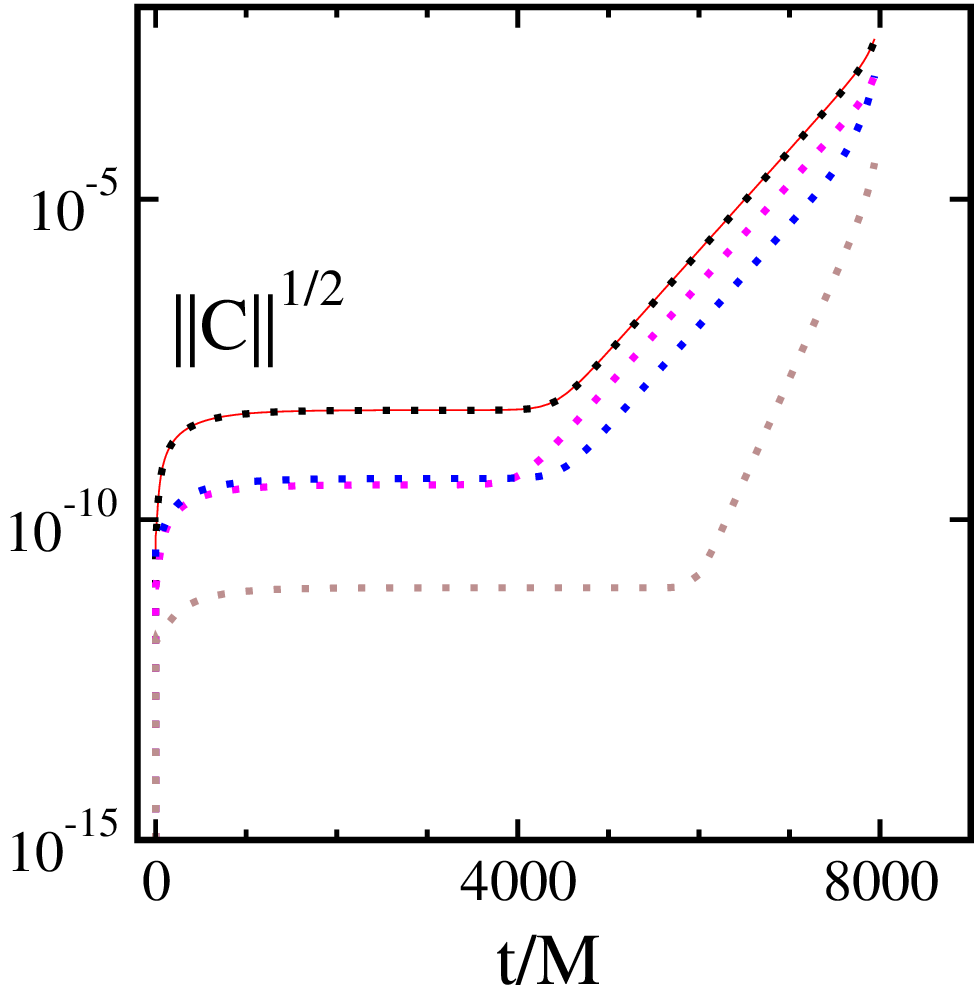}
\end{center}
\caption{Solid curve shows the evolution of the
integral norm of all the constraints $||{\cal C}_{kij}{\cal
C}^{kij}+{\cal C}_k{\cal C}^k +{\cal C}_{klij}{\cal C}^{klij}+{\cal
C}^2||^{1/2}$ (per unit volume) for the most stable
set of evolution parameters.  Dotted curves show the individual
contributions from the various constraints: ${\cal C}_{kij}{\cal
C}^{kij}$, ${\cal C}_{klij}{\cal C}^{klij}$, ${\cal C}_k{\cal C}^k$
and ${\cal C}^2$, (in that order from largest to smallest at late times). }
\label{f:Constraints}
\end{figure}

Figure~\ref{f:Enorms} shows numerical results from the evolution of a
single black hole for the case $\gamma=-12$, $\zhat=-0.425$.  Plotted
is the energy norm $||\delta E||{}^{1/2}$ (as introduced in
Ref.~\cite{Lindblom2002}), which measures the deviation of the
numerical solution from an exact solution that satisfies the
constraints. The solid curves in Figure~\ref{f:Enorms} represent
computations performed at different spectral resolutions (18, 24, and
32 radial collocation points), and thus illustrate the convergence of
our solutions. The dashed curve represents the evolution obtained with
a linearized version of the code, normalized so that the amplitude of
the unstable mode is the same as that obtained with the non-linear
evolution.  The convergence of these solutions, as illustrated in
Fig.~\ref{f:Enorms}, is made possible by choosing the same initial
data, including the exact same form for the initial perturbation
added by hand to Eq.~\eqref{e:pgmetric}, for each resolution. If we had
instead chosen initial data given by Eq.~\eqref{e:pgmetric} plus {\it
random\/} perturbations (either supplied by numerical roundoff error
or introduced by hand) we would not expect results using different
resolutions to converge to the same solution.

Figure~\ref{f:Constraints} shows the evolution of the integral
norm of the constraints (see Refs.~\cite{Kidder2001,Lindblom2002} for
definitions of the constraint variables) for the highest-resolution
case shown in Figure~\ref{f:Enorms}.  Note that at late times, most of
the constraints in Figure~\ref{f:Constraints} grow at the same rate ($1/\tau
\approx 1/275M$) as the energy norm shown in Figure~\ref{f:Enorms}. The
exception is the Hamiltonian constraint, which is much smaller than
the other constraints, but grows at double the growth rate,
$1/\tau\approx 1/137M$. Thus it appears that for the optimal choice of
parameters, the unstable mode violates the Hamiltonian constraint only
to second order in the mode amplitude.

\begin{figure}
\begin{center}
\includegraphics[width=3.0in]{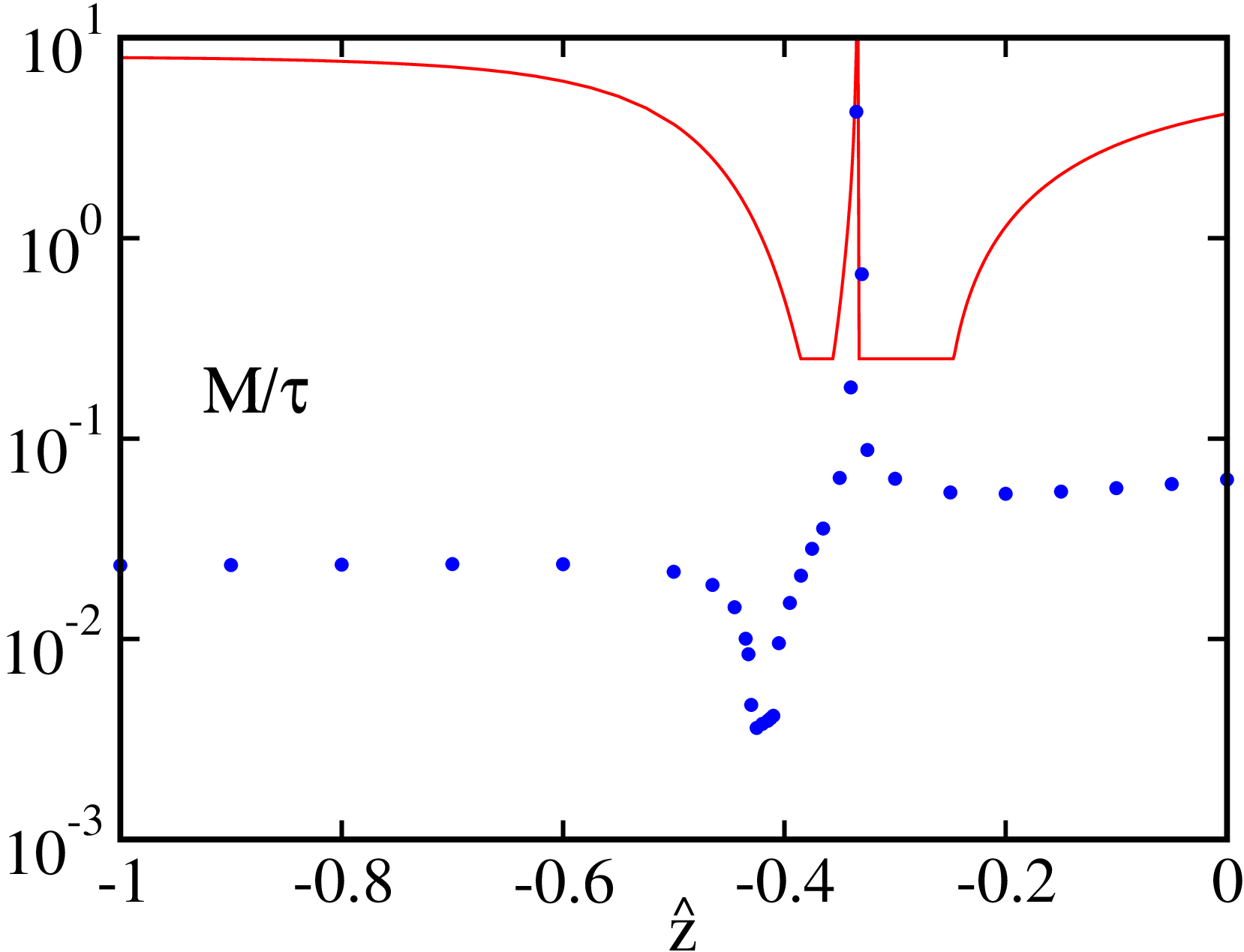}
\end{center}
\caption{Exponential growth rates of the constraint-violating
instabilities as a function of the parameter $\zhat$ (with fixed
$\gamma=-12$).  Points are numerically
determined rates, while the solid curve is the approximate
growth rate.}
\label{f:ZhatDependence}
\end{figure}

\begin{figure}
\begin{center}
\includegraphics[width=3.0in]{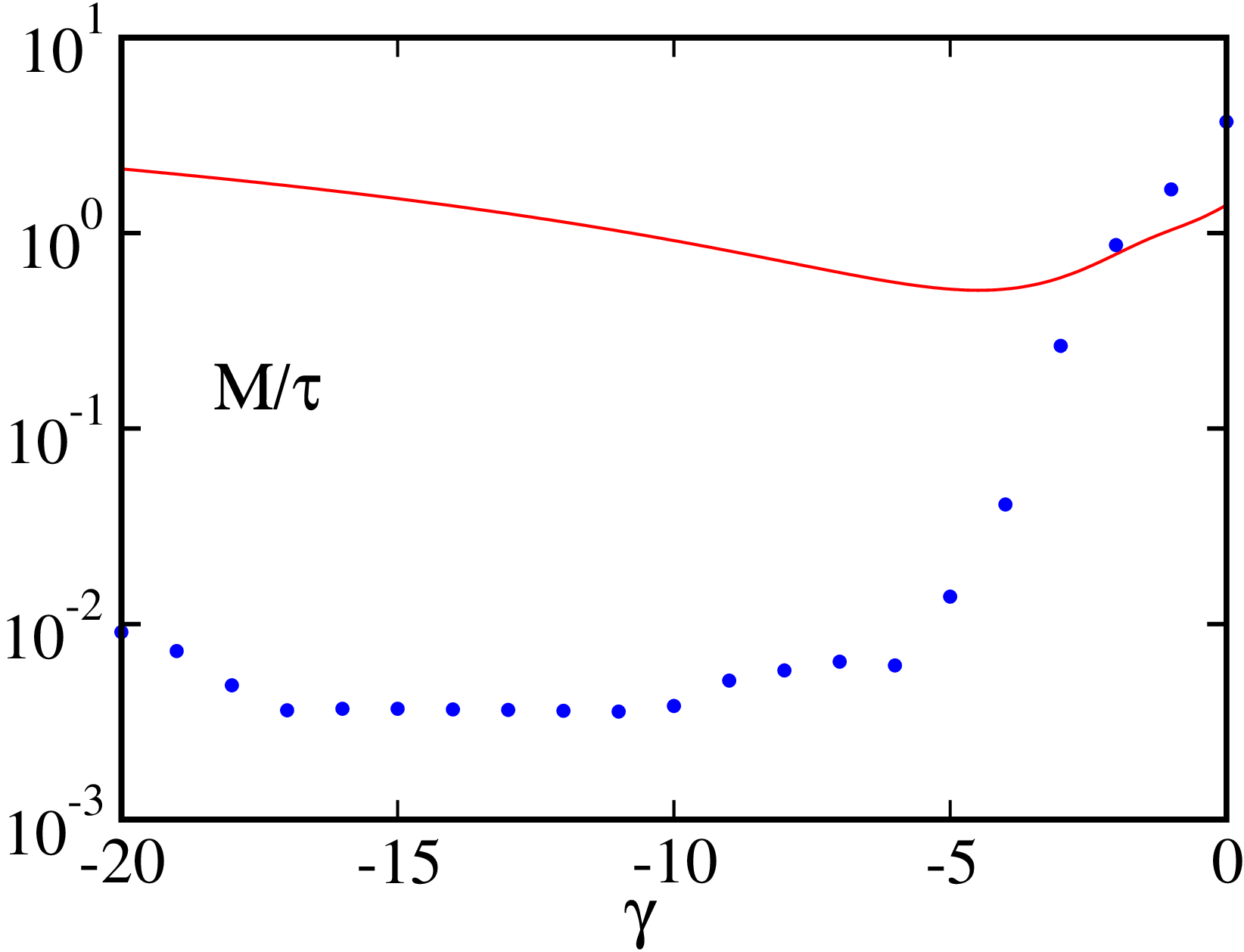}
\end{center}
\caption{Exponential growth rates of the constraint-violating
instabilities as a function of the parameter $\gamma$ (with fixed
$\zhat=-0.425$).  Points are numerically
determined rates, while the solid curve is the approximate
growth rate.}
\label{f:GammaDependence}
\end{figure}

\begin{figure}
\begin{center}
\includegraphics[width=3.0in]{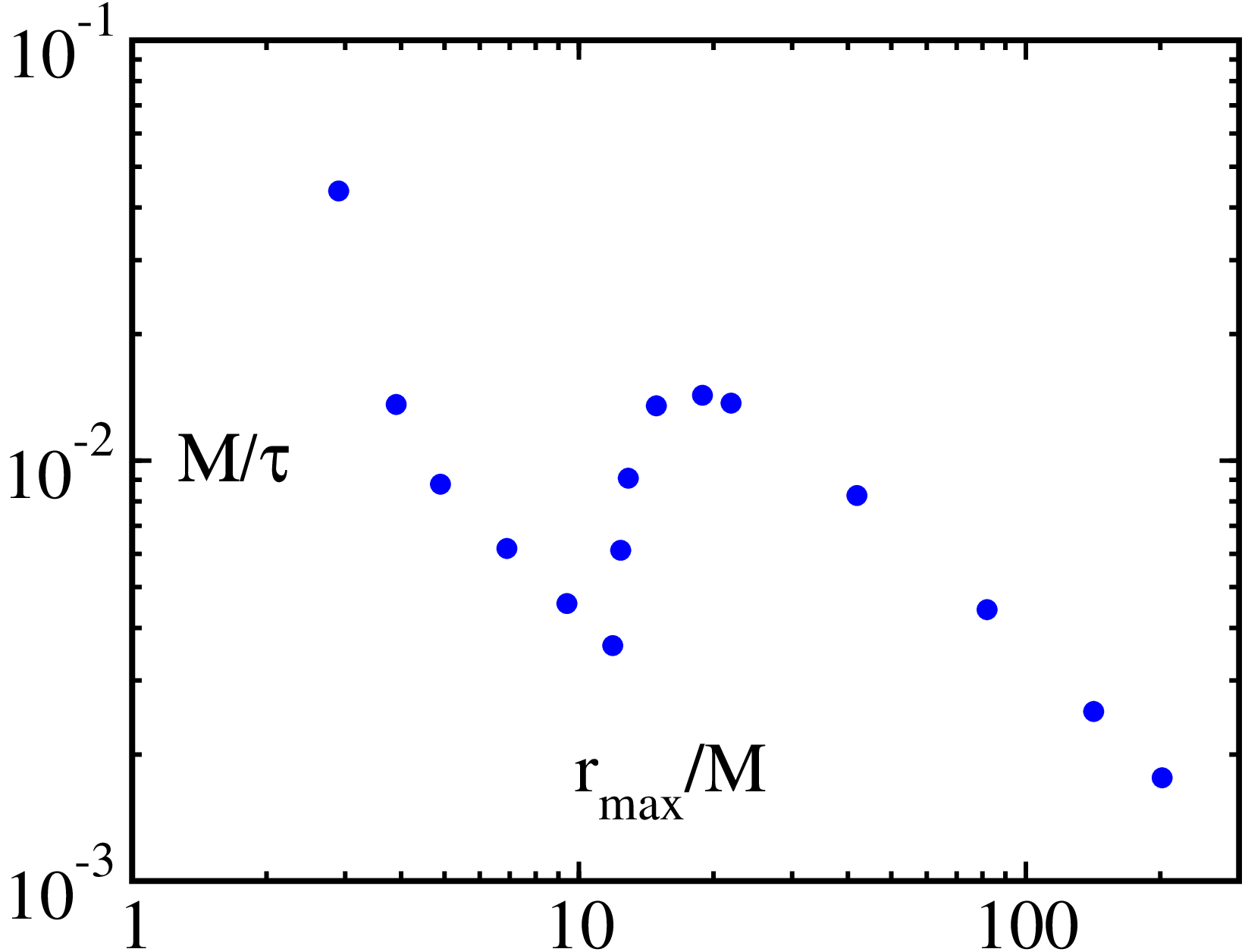}
\end{center}
\caption{Instability growth rates as a function of the location of the
outer boundary of the computational domain for the evolution parameter
values $\gamma=-12$, $\zhat=-0.425$.}
\label{f:OuterBoundary}
\end{figure}

Given a numerical evolution for a particular set of parameters, we
determine the exponential growth rate by measuring the slope of the
curve in Figs.~\ref{f:Enorms} or~\ref{f:Constraints}. Figures~\ref{f:ZhatDependence} and
\ref{f:GammaDependence} illustrate these growth rates as functions of the
parameters $\gamma$ and $\zhat$. The points in these figures represent
numerically determined growth rates measured using the linearized code
(which yields the same growth rates as the fully
nonlinear code, see Figure~\ref{f:Enorms} and
Ref.~\cite{Lindblom2002}). The solid curves represent the simple {\it
a priori} estimates of these growth rates introduced in
Ref.~\cite{Lindblom2002}. Although the agreement between the estimates
and the numerical results is only approximate, this agreement was good
enough to allow us to direct our search for the most stable values of
the parameters to the relevant region of the parameter space.  The
curves in these figures represent orthogonal slices of the function
$1/\tau(\gamma,\zhat)$ through its minimum, $1/\tau_{\max}=1/275M$,
which occurs at the parameter values $\gamma=-12$ and $\zhat=-0.425$.
This minimum growth rate is such that constraint violations in the
initial data that are comparable to typical machine precision
(e.g. $10^{-16}$) will become large (e.g. of order 0.1) when $t\approx
10^4M$.  Figures~\ref{f:Enorms} and~\ref{f:Constraints} illustrate the full
non-linear evolution that corresponds to this optimal choice of
parameters.

For all the cases discussed so far, the outer boundary radius was set
at $r_{\rm max}=11.9M$.  Figure~\ref{f:OuterBoundary} illustrates the
dependence of the growth rate $1/\tau$ on the location of the outer
boundary of our computational domain, for fixed $\gamma=-12$ and
$\zhat=-0.425$.  This curve shows a sharp local minimum at the radius
where the optimal set of evolution parameters $\{\gamma,\zhat\}$ was
determined, strongly suggesting that these optimal values depend on
the location of this outer boundary.  We have verified this by
studying in some detail the case where the outer boundary is located
at $r_{\rm max}=81.9M$.  There we find that the new optimal values of
the parameters become $\gamma=-12$ and $\zhat=-0.41$, and the value of
the growth rate at these new optimal parameters becomes $1/\tau =
1/333M$.  This growth rate is about 20\% smaller than that of the
system whose evolution is illustrated in Fig.~\ref{f:Enorms}. Thus we
infer that the evolution of a single black hole in this case would
extend to about $10^4M$.  We also note that the optimal parameters
for $r_{\rm max}=81.9M$ give a value of $1/\tau$ that is about 2/3 the
value illustrated in Fig.~\ref{f:OuterBoundary} for this value of
$r_{\rm max}$.  Considerable additional computational effort will be
required to determine the general dependence of the optimal value of
$1/\tau$ on $r_{\rm max}$, and we postpone that to a future study.
For $r_{\rm max}<12M$ and for $r_{\rm max}>20M$, the growth rate for a
fixed set of evolution parameters decreases roughly like
$\Lambda/r_{\rm max}$, with the constant $\Lambda$ being about a
factor of six larger for the case with $r_{\rm max}>20M$.  However,
the {\it optimal\/} value of $1/\tau$ as a function of $r_{\rm max}$
does not scale in this simple way.  The smallest growth rate
determined in our study to date is the point at $\gamma=-12$ and
$\zhat=-0.425$ with $r_{\rm max}=201.9M$, where we find
$1/\tau=1/570M$.  This evolution would be expected to persist for over
16,000$M$.

Finally, we note that all of the numerical evolutions discussed so far
have placed the {\it inner\/} boundary of the domain at $r_{\rm
min}=1.9M$.  We have also run the code with $r_{\rm min}=1.0M$ and
$r_{\rm min}=1.5M$ for our best-studied case
$(\gamma=-12,\zhat=-0.425,r_{\rm max}=11.9M)$ and we find that the
growth rate is the same to three significant digits.

In summary, we have illustrated that significant improvements in the
stability of numerical evolutions of 3D black-hole spacetimes can be
achieved by a careful choice of the representation of the Einstein
evolution equations.  In particular we have shown that single black
hole spacetimes can be evolved longer than $t\approx 8000M$
even with fixed gauge.  These new results also indicate that the outer
boundary conditions may play a significant role in fixing the optimal
formulation of the equations, as has been suggested by other
investigations~\cite{FriedrichNagy1999,%
Calabrese2001,Szilagyi2002,Calabrese2002c}.
The role of these boundary conditions will be explored more thoroughly
in a future study.

\acknowledgments Some computations were performed on the IA-32 Linux
cluster at NCSA. This research was supported in part by NSF grant
PHY-0099568 and NASA grant NAG5-10707 at California Institute of
Technology and NSF grants PHY-9800737 and PHY-9900672 at Cornell
University.

\end{document}